\newif\ifSYNASC\SYNASCtrue

\def\N{\mathbf{ N}}

\documentclass[conference]{IEEEtran}
\IEEEoverridecommandlockouts
\usepackage[hyphens]{url}
\usepackage{pdfpages}
\usepackage{verbatim}
\usepackage{hyperref}
\usepackage{combelow}
\usepackage[show]{ed}
\usepackage{graphicx}
\usepackage{balance}

\begin{document}
\author{\IEEEauthorblockN{
James H. Davenport}
\IEEEauthorblockA{\textit{Departments of Computer Science and Mathematical Sciences} \\
\textit{University of Bath}\\
Bath BA2 7AY, United Kingdom \\
masjhd@bath.ac.uk \quad  0000-0002-3982-7545}}
\title{First steps towards Computational Polynomials in Lean\\
\thanks{The author is partially supported by EPSRC under grant EP/T015713/1. This research started at the Hausdorff Institute supported by  Deutsche Forschungsgemeinschaft (DFG, German Research Foundation) under Germany's Excellence Strategy – EXC-2047/1 – 390685813. Several participants at \url{https://www.mathematics.uni-bonn.de/him/programs/current-trimester-program/him-trimester-program-prospects-of-formal-mathematics} have given useful suggestions, and Mario Carneiro has been a constant source of advice and implementation. The author is grateful to the SYNASC referees, James McKinna and Ali Uncu for comments on drafts.}}
\maketitle
\def\x{{\bf x}}
\begin{abstract}
	The proof assistant Lean has support for abstract polynomials, but this is not necessarily the same as support for computations with polynomials. Lean is also a functional programming language, so it should be possible to implement computational polynomials in Lean. It turns out not to be as easy as the naive author thought.
\end{abstract}
This is based on the author's experience in computer algebra, and \cite{Davenport2022b}. We consider polynomials in commuting variables over a commutative ring $R$ (non-commutativity is considered in \cite[\S2.4]{Davenport2022b}). To avoid a plethora of ``\dots'', we will tend to use three variables $x,y,z$ as our example, or $\x$ if we want a vector of variables.
\section{State of Lean}
\subsection{Existing support}
\def\foo#1{\ifSYNASC\item[$\bullet$]{\bf #1. }\else\item[#1]\fi}
\begin{description}
\foo{Univariate}See \url{https://leanprover-community.github.io/mathlib4_docs/Mathlib/Algebra/Polynomial/Basic.html}.
\foo{Multivariate}See \url{https://leanprover-community.github.io/mathlib4_docs/Mathlib/Algebra/MvPolynomial/Basic.html}. This ``creates the type \verb+MvPolynomial+ $\sigma$ $R$, which mathematicians might denote $R[X_i: i\in\sigma]$''. 
\item[Note]That $\sigma$ might be infinite, whereas in the rest of the document, by ``multivariate'' we mean ``in a fixed set of variables''.
\end{description}
\subsection{Why both?}
The reader might ask: ``Isn't univariate just multivariate with $\sigma$ being a singleton?'' In one sense this is true, but not very helpful. If I have two univariate polynomials over a field $f:=x^a+\cdots$ and $g:=x^b+\cdots$, then I can divide one by the other (which way round depends on whether $a>b$) to get a remainder $h:=x^c+\cdots$ with $c<\min(a,b)$.  In Axiom-speak, the exponents $\N$ are an \verb+OrderedCancellationAbelianMonoid+, i.e. $a\ge b\rightarrow\exists c: a=b+c$.  This is not true of multivariates: consider $f=x^1y^0$ and $g=x^0y^1$.
If the polynomials are not over a field, but merely an integral domains, we can still do pseudo-division of univariates.
Subject to the appropriate conditions on the ground field, this means that univariates over a field form a Euclidean domain, whereas multivariates do not.
\subsection{A curious case}
Lean rings include the ring with one element, so that $0=1$. Hence polynomials over this ring are trivial, as $x$ can't have a non-zero coefficient. This creates various special cases. For the moment we allow this case, though JHD does wonder whether it is introducing too much inefficiency. This will need to be checked, but fortunately Lean has good profiling tools \cite{Ullrich2024a}.
\section{Representations of Polynomials}
We have various decisions to make over the computational representation of polynomials.
\subsection{A preliminary choice}
Do we store zero terms?
\begin{description}
\item[Dense]All terms in $f=\sum_{i=0}^n a_ix_i$ from $i=0$ to $i=n$ are stored, whether or not $a_i=0$. Well-suited to a vector representation.
\item[Sparse]Only store the terms with $a_i\ne 0$, but need to store $i$ with $a_i$ to tell $x^2+2$ from $x^2+2x$.
\end{description}
General-purpose representation in computer algebra systems are always sparse, otherwise you look stupid if you can't handle $x^{1000000000}+1$
\subsection{Geobuckets}\label{sec;Geo}
The obvious way to implement a sparse data structure is as a list of pairs $(i,a_i)$, sorted according to $i$ and with duplicates combined. This is very close to the pen-and-paper representation. The cost of adding polynomials with $m$ and $n$ terms is $m+n-1$ comparisons. But note that the \emph{cost} of addition is not associative: if $p,q,r$ are polynomials with $l,m,n$ terms respectively, and no combination of terms happens, then $p+(q+r)$ takes $l+2(m+n)-2$ comparisons while $(p+q)+r$ takes $2(l+m)+n-2$ comparisons. This shows up in Gr\"obner basis computations, where we are typically reducing a long polynomial $p$ by the generators, which tend to be much shorter. This asymmetry is addressed by the geobucket data structure \cite{Yan1998}, which has been adopted by many specialist systems \cite{Abbott2015a,Schonemann2015a}. 

A polynomial is stored as an (unevaluated) sum of polynomials, with the $k$th polynomial having at most $c^k$ terms (typically $c=4$) --- hence \emph{geo}metrically increasing \emph{buckets}. If we add a regular polynomial with $\ell$ terms to a geobucket, we add it to bucket $k$ with $c^{k-1}<\ell\le c^k$, and if the result has more than $c^k$ terms, we add that to bucket $k+1$, and cascade the overflow as necessary. A further advantage of the geobucket structure is given in \S\ref{sec:Read}.
\subsection{An improvement}
\cite[Figure 2]{Yan1998} suggests looking for the leading coefficient among the leading coefficients of the buckets. At least one of \cite{Abbott2015a,Schonemann2015a} told the author that instead they ensured that the leading coefficient was always the leading coefficient of the largest bucket.
\subsection{Related work}
The geobucket structure can be seen as deferring the expensive addition of differently-sized objects until the last possible moment. In this sense it has similarities to the Hierarchical Representation of \cite{Zhouetal2006}, as improved in \cite{Stoccoetal2024a}. The principal difference is that in the hierarchical representation the user decides when to ``unveil'' (i.e. use the internal structure) an object, whereas in geobuckets the system automatically unveils the appropriately sized bucket, and actively updates the buckets. Hence the two systems have different utilities.
\section{Choices of Multivariate Polynomial Representations}
Mathematically, $R[x,y,z]$ is the same structure as $R[x][y][z]$ (were it not for \cite{Buzzard2024a}, the author would have written $R[x,y,z]=R[x][y][z]$). 
When it comes to computer representations, this leads to a major choice.
\begin{description}
\foo{Distributed}This is $R[x,y,z]$, and is the representation of choice for Gr\"obner base algorithms. We will normally fix in advance our set of variables, and a total order $\prec$ on the monomials, which tells us whether $x^\alpha y^\beta z^\gamma\prec x^{\alpha'} y^{\beta'} z^{\gamma'}$.  It is necessary in Gr\"obner base theory, and helpful in implementation, to assume that $\prec$ is compatible with multiplication: {\def\i{{\bf i}}
\def\j{{\bf j}}\def\k{{\bf k}} $\x^\i\prec\x^\j\Rightarrow \x^{\i+\k}\prec\x^{\j+\k}$}.
\par
If we have $k$ variables, then the obvious technique is to store the term $cx^\alpha y^\beta z^\gamma$ as $(\alpha,\beta,\gamma,c)$ (or possibly a record structure. However, since we often use total degree orders in Gr\"obner base computation, the Axiom implementation\footnote{This is now also done in Maple: \cite{MonaganPearce2014b}.} actually stored $(\alpha+\beta+\gamma,\alpha,\beta,\gamma,c)$, i.e. the total degree first.
\foo{Recursive}A typical representation for, say, $x^3-2x$ would be $(x,(3,1),(1,-2))$, i.e. a list starting with the variable, then ordered pairs as in ``Sparse'' above. 
In a typed language we might have a record type \verb![variable,list]!. 
Hence $z^2(y^2+2)+(3y+4)\in R[y][z]$ would be represented as \begin{equation}(z,(2,(y,(2,1),(0,2))),(0,(y,(1,3),(0,4)))).\end{equation}
	What about $z^2(y^2+2)+(3x+4)\in R[x][y][z]$?
		There are (at least) two options.
\foo{Dense in variables}In this option it would be represented as
\ifSYNASC
		\begin{equation}\begin{array}{c}(z,(2,(y,(2,(x,(0,1))),(0,(x,(0,2))))),\\(0,(y,(0,(x,(1,3),(0,4)))))).\end{array}\end{equation}
			\else
		\begin{equation}(z,(2,(y,(2,(x,(0,1))),(0,(x,(0,2))))),(0,(y,(0,(x,(1,3),(0,4)))))).\end{equation}
			\fi
\foo{Sparse in variables}In this option it would be represented as \begin{equation}(z,(2,(y,(2,1),(0,2))),(0,(x,(1,3),(0,4)))).\end{equation}
\item[So]``Sparse in variables'' would seem easier, but the snag is that we can meet two polynomials with different main variables, and we need some way of deciding which is the `main' variable, else we can end up with polynomials in $y$ whose coefficients are polynomials in $x$ whose coefficients are polynomials in $y$, which is not well-formed.
\foo{Other}There are other options, with interesting complexity-theoretic implications, but not used in mainstream computer algebra: see \cite[\S2.1.5]{Davenport2022b}.
\end{description}
Experience in Axiom, as in \cite{Davenportetal1991a}, shows that it may be useful to be able to talk about univariate polynomials in an unspecified variable, i.e. just a list of (exponent,coefficient) pairs with no variable specified.
\par
Recursive is suited to algorithms such as g.c.d., factorisation, integration etc., in fact almost everything except Gr\"obner bases. Most systems therefore use recursive, and for example, when implementing Gr\"obner bases in Reduce, which is recursive, \cite{GebauerMoller1988} implemented their own distributed polynomials, and converted in/out.
\subsection{Addition of Sparse polynomials}\label{sec:add}
This should be simple: merge the lists except when both have the same exponent, when we add the coefficients (and handle the case where the sum is zero specially). This should translate into Lean like this (in fact for the multivariate case).
\ifSYNASC The code is shown in Figure \ref{LeanAdd}, where \verb+nvars+ is the number of variables and \verb+MvDegrees+ can be thought of\footnote{In practice it also stores the total degree, to make total degree order comparisons faster.} as the \verb+nvars+-tuple of degrees in these variables.\begin{figure*}[t]\caption{Lean addition\label{LeanAdd}}\fi
\begin{verbatim}
def addCore : List (MvDegrees nvars × R) → List (MvDegrees nvars × R) 
                      → List (MvDegrees nvars × R)
  | [], yy => yy
  | xx, [] => xx
  | xx@((i, a) :: x), yy@((j, b) :: y) =>
    if i < j then
      (j, b) :: addCore xx y
    else if j < i then
      (i, a) :: addCore x yy
    else  -- check for a+b=0
      ( fun c => if c=0 then addCore x y else (i, c) :: addCore x y) (a+b)
\end{verbatim}
	The notation \verb!xx@((i, a) :: x)! means ``call it \verb+xx+, but also deconstruct it into \verb+(i, a)+ as the head, and \verb+x+ as the tail. 
\ifSYNASC\end{figure*}
\par
Note that Lean requires a recursive definition to be proven to terminate.
Any programmer would say that termination is obvious, as every recursive call is on less (either less \verb+x+ or less \verb+y+, or possibly both), but Lean doesn't recognise this, as it says below (where we have used ``R'' for ``recursion''). 
\begin{verbatim}
fail to show termination for
  MvSparsePoly.addCore
with errors
argument #5 was not used for structural R
  failed to eliminate recursive application
    addCore xx y

argument #6 was not used for structural R
  failed to eliminate recursive application
    addCore x yy

structural R cannot be used
\end{verbatim}
\cite{McKinna2024a} reports that an ``obvious'' translation into Agda has a similar issue with automatic proof of termination, and it needs a ``non-obvious'' translation to allow the automatic termination prover to conclude termination.
Proofs of termination in Agda are discussed in \cite{Agda2024a}.
\par
The author's Lean solution is
\begin{verbatim}
termination_by xx yy =>
    xx.length + yy.length
\end{verbatim}
which requires the \verb+xx@+ syntax to ensure that \verb+xx+ was defined.
\par
\begin{figure*}
	\caption{Definition of a multivariate polynomial\label{Fig:MvPoly}}
	\hbox{\hskip-10pt\includegraphics{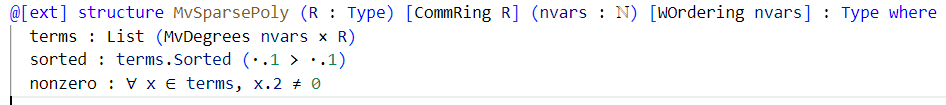}}
\end{figure*}
\begin{figure*}
	\caption{Sorted addition of multivariate polynomials\label{Fig:AddCoreSorted}}
	\hbox{\hskip-10pt\includegraphics{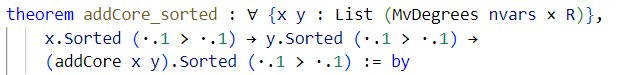}}
\end{figure*}
We need more than this, though. A true polynomial is defined in Figure \ref{Fig:MvPoly} (where \verb+Wordering+ is the type for the well-ordering on monomials), and in particular we require the list of terms to be sorted. This requires a theorem as in Figure \ref{Fig:AddCoreSorted}. We are currently working on a proof of this.
\subsection{Polynomial Multiplication}
Naive polynomial multiplication (but see \S\ref{sec:Mult}) is relatively simple to implement given addition. Similarly the proofs that this algorithm gives well-formed results should follow from those results for addition. We have yet to see what the challenge is for correctness.
\section{Polynomial Verification}
\cite{Buzzard2024b}  states that, asked to verify $f\in\langle f_1,\ldots,f_k\rangle$, Lean asked Sage (actually Singular) to compute the co-factors $\lambda_i$, and Lean just verifies that $f=\sum^N\lambda_if_i$. This verification isn't really a Gr\"obner basis computation, and any polynomial representation would be appropriate. Which is best? Experiments will need to be performed, but an obvious guess would be the same representation as Sage outputs. 
\par
Let $\#f$ denote the number of terms in the polynomial $f$.
\subsection{A specialist input mechanism?}\label{sec:Read}
Note that, at least in the worst case, the $\lambda_i$ may be much larger than $f$ or $f_i$.
For example, in proving non-singularity of Weierstrass-form elliptic curves \cite[\S3.2]{Macbeth2024b}, the largest $f_i$ is $-y^2+2pxz+3qz^2i:=f_3$, but the largest $\lambda_i$ is
\[\begin{array}{c}
(256 / 3)  p ^ {12}  x ^ 2 + 128  q  p ^ {11}  x  z + 2304  q ^ 2  p ^ 9  x ^ 2 +
                                2592  q ^ 3  p ^ 8  x  z -\\
			      64  q  p ^ {10}  y ^ 2 +
                            23328  q ^ 4  p ^ 6  x ^ 2 +
                          17496  q ^ 5  p ^ 5  x  z -
                        1296  q ^ 3  p ^ 7  y ^ 2 +\\
                      104976  q ^ 6  p ^ 3  x ^ 2 +
                    39366  q ^ 7  p ^ 2  x  z -
                  8748  q ^ 5  p ^ 4  y ^ 2 +\\
                177147  q ^ 8  x ^ 2 -
              19683 q ^ 7 p y ^ 2  =:\lambda_1
\end{array}\]
with 13 terms, and in fact $f_1\lambda_1$ has 26 terms.

If the $\lambda_i$ are output in decreasing order, then the naive algorithm ``read a term; add it to the polynomial'' using the method of \S\ref{sec:add}, will take $O((\#\lambda_i)^2)$ operations. A geobucket implementation  (\S\ref{sec;Geo}) would make this $O(\#\lambda_i\log(\#\lambda_i))$. However, an ``intelligent'' algorithm that knew it was reading in a specific order would be simply  $O(\#\lambda_i)$.
\subsection{\cite{Johnson1974}, as described in \cite{MonaganPearce2009b}}\label{sec:Mult}
This is an algorithm for sparse polynomial multiplication. Let $f$ and $g$ be polynomials stored in a sparse 
format that is sorted with respect to a monomial order $\prec$.
We shall write the terms of $f$ as $f = f_1 + f_2 + \cdots + f_{\#f}$ and
$g$ as $g = g_1 + g_2 +\cdots + g_{\#g}$. We require $g$ to be stored as a list, so that $g-g_1$ (and then $g-g_1-g_2$ etc.) can be computed on $O(1)$ time and 0 space (i.e. \verb+CDR+ in Lisp). Our task is to compute the
product $h = f \times g =\sum_{i=1}^{\#f}\sum_{j=1}^{\#g}f_ig_j$ term-by-term, starting with the greatest. 
\par
We use a (balanced) binary heap to merge each (unevaluated) $f_i\times g$, so the sort key is $\deg(f_i)+\deg(g)$, and the heap has $\#f$ entries. After an $f_i\times g_j$ is extracted, we insert $f_i\times (g_{j+1}+\cdots)$. If the leading monomial of the resulting heap is the same as the just-extracted  $f_i\times g_j$, we add the coefficients and continue.  Since the size of the heap is $\#f$, the cost of rebalancing after a single extraction is $\log\#f$, the number of extractions is $\#f\#g$, and therefore the total cost is $\#f\#g\log\#f$.
\par
We could therefore solve the problem from \cite{Buzzard2024b} by computing each $\lambda_if_i$ this way, with $\lambda_i$ being $g_i$ as we might expect $\#\lambda_i>\#f_i$ at least in the worst case. This would still requiring storing each $\lambda_if_i$ (but if we computed  $\sum\lambda_if_i$ as we went along, we would store the partial sum and the single $\lambda_if_i$ being added).
\subsection{Verification without product construction}
The challenge is to compute $\sum^nf_i\lambda_i$ efficiently. In particular $f_i\lambda_i$ may well be much larger than either $f_i$ or $\lambda_i$
As in \cite{Johnson1974}, represent each $\lambda_if_i$ (unevaluated) by a binary heap $H_i$. Now create a binary heap $H$ whose elements are the heaps $H_i$. Extract the leading coefficient of $H$ (as above, this may be the sum of several such, and indeed, because we are expecting cancellation in $\sum^N\lambda_if_i$, it may well be zero).  $H$ has $N$ elements, so the overall cost of this operation is $\sum^N\#\lambda_i\#f_i(\log N+\log \#f_i)$. Note that there is no $\log\#\lambda_i$ term --- this is because we are leveraging the fact that the $\lambda_i$ are already sorted.
\par
Write $f=\sum^N f_i\lambda_i$. Then we certainly need to store\footnote{We should note that different choices of $\lambda_i$ can have different storage requirements: see \cite{HofstadlerVerron2024a}.} the $\lambda_i$. We need to store the $f_i$, which are stored in the heap structures $H_i$, taking space $O(\sum\#f_i)$ (probably with a higher constant of proportionality). There's also $O(N)$ for the master heap $H$.  There's also the cost of the output.  Note that the output $f$ is constructed term-by-term, and no term is deleted, so the space cost is $O(\#f+\sum_N\#f_i)$.
\par
A further improvement, rather than verifying  $f=\sum^N f_i\lambda_i$, is to verify $0= (-1)f+\sum^N f_i\lambda_i$, so there is no output to construct.
It would also be possible to run the heaps with minimal degree at the root (i.e. verifying the trailing coefficient first), which might be more advantageous in the case where there is actually an error, i.e.  $0\ne (-1)f+\sum^N f_i\lambda_i$. 
\subsection{Interaction with geobuckets}
\cite{Johnson1974} assumed that the $g_i$ are stored as lists: what if they are stored as geobuckets? Then the process of replacing $g$ by $g-g_1$ is no longer $O(1)$, but depends on how the overall geobucket structure is stored: if it's a vector of $b$ buckets, then the cost is $O(b)=O(\log \#g_i)$. There are (at least) three possible routes.
\begin{enumerate}
	\item Do an initial conversion of $g$ to list structure, adding the buckets from small to large, and then using \cite{Johnson1974}. The cost is $\#g$ in time (and space!).
	\item Instead of storing $f_ig$ in the  \cite{Johnson1974} heap, store each $f_ig^{(j)}$ where the $g^{(j)}$ are the buckets of $g$. This means there are $b$ times as many entries in the heap, so multiplies the cost of  \cite{Johnson1974}  by $O(\log b)=O(\log\log\#g)$
	\item In practice we would probably do a hybrid: add the small buckets together but store $f_ig^{(j)}$ separately for large $g^{(j)}$, where ``large'' is a tuning parameter.
\end{enumerate}

\bibliography{../../../jhd}
\end{document}